\documentclass{llncs}

\usepackage[dvips]{graphicx}
\usepackage{subfigure}
\usepackage{psfrag}
\graphicspath{fig} 

\DeclareGraphicsExtensions{.ps,.eps,.ps.gz,.eps.gz,.ps.Z,.eps.Z}

\newcommand{\norm}[1]{ \left\Vert {#1} \right\Vert}

\newcommand{\beq}{\begin{equation}}
\newcommand{\beqa}{\begin{eqnarray*}}
\newcommand{\beqan}{\begin{eqnarray}}
\newcommand{\greq}{\begin{equation}\left\{ \begin{array}{l}}
\newcommand{\egreq}{\end{array}\right. \end{equation}}
\newcommand{\nngreq}{\[\left\{ \begin{array}{l}}
\newcommand{\nnegreq}{\end{array}\right. \]}
\newcommand{\egreqn}[1]{\end{array}\right. \label{#1}\end{equation}}
\newcommand{\eeq}{\end{equation}} 
\newcommand{\eeqn}[1]{\label{#1}\end{equation}} 
\newcommand{\eeqa}{\end{eqnarray*}}
\newcommand{\eeqan}[1]{\label{#1}\end{eqnarray}}

\newcommand{\hpp}{ \hspace{2pt} }
\newcommand{\noi}{ \noindent }

\newcommand{\lp}{ \left(}
\newcommand{\rp}{ \right)}

\newcommand{\ulm}{u^\ell_m}
\newcommand{\vlm}{v^\ell_m}
\newcommand{\wlm}{w^\ell_m}

\newcommand{\YL}{ Y^m_\ell }

\newcommand{\RL}{ \vec{R}^m_\ell }
\newcommand{\SL}{ \vec{S}^m_\ell }
\newcommand{\TL}{ \vec{T}^m_\ell }

\newcommand{\iml}{ \frac{im}{\ell(\ell+1)} }

\newcommand{\na}{ \vec{\nabla} }

\newcommand{\vu}{\vec{u}}

\newcommand{\ez}{\vec{e}_z}

\newcommand{\dr}[1]{\frac{\partial  #1}{\partial r}}

\newcommand{\dnr}[1]{\frac{d  #1}{dr}}

\newcommand{\ddnr}[1]{\frac{d^2  #1}{dr^2}}

\def\Div{\mathop{\hbox{div}}\nolimits}

\renewcommand{\Div}{\vec{\nabla}\cdot}
\newcommand{\disp}[1]{\displaystyle #1}
\newcommand{\eq}[1]{(\ref{#1})}

\def\og{\leavemode\raise.3ex\hbox{$\scriptscriptstyle\langle\!\langle$}}
\def\fg{\leavemode\raise.3ex\hbox{$\scriptscriptstyle\rangle\!\rangle$}}

\newtheorem{algorithm}{Algorithm}

\begin{document}
\bibliographystyle{plain}

\mainmatter              
\title{Convergence and round-off errors in a 
two-dimensional eigenvalue problem using spectral methods and Arnoldi-chebyshev
algorithm
}
\titlerunning{Convergence and round-off errors in a  
two-dimensional eigenvalue problem}
\author{Lorenzo Valdettaro\inst{1} \and Michel Rieutord\inst{2}
\and Thierry Braconnier\inst{3} \and Val\'erie Frayss\'e\inst{4}
}
\authorrunning{Lorenzo Valdettaro et al.}   

\institute{Dipartimento di Matematica, Politecnico di Milano, 
Piazza L. da Vinci 32,\\ 
I-20133 Milano, Italy\\
\and
Observatoire Midi-Pyr\'en\'ees,
14 av. E. Belin,
 F-31400 Toulouse, France
\and
Department of Mathematics, University of Manchester,
  Manchester, M13 9PL, UK.
\and
CERFACS, 42, Avenue Coriolis, F-31057 Toulouse, France
}

\maketitle              

\begin{abstract}
An efficient way of solving 2D stability problems in fluid mechanics is
to use, after discretization of the equations that cast the problem in the
form of a generalized eigenvalue problem, the incomplete Arnoldi-Chebyshev
method. This method preserves the banded structure sparsity of matrices
of the algebraic eigenvalue problem and thus decreases memory use and
CPU-time consumption.

The errors that affect computed eigenvalues and eigenvectors are due
to the truncation in the discretization and to finite precision in the
computation of the discretized problem.  In this paper we analyze those
two errors and the interplay between them.  We use as a test case the
two-dimensional eigenvalue problem yielded by the computation of inertial
modes in a spherical shell.  This problem contains many difficulties
that make it a very good test case.  It turns out that that single
modes (especially most-damped modes i.e. with high spatial frequency)
can be very sensitive to round-off errors, even when apparently good
spectral convergence is achieved.  The influence of round-off errors is
analyzed using the spectral portrait technique and by comparison of double
precision and extended precision computations.  Through the analysis we
give practical recipes to control the truncation and round-off errors
on eigenvalues and eigenvectors.

\end{abstract}

\section{Introduction}

The first step in studying the stability of the solutions of a nonlinear
problem, is to solve the eigenvalue problem associated with
infinitesimal pertubations which are superposed to the equilibrium state.
Even if the equations of these perturbations are linear, solving
the eigenvalue problem may be a formidable task. The difficulties arise
in general when the variables do not separate. In such a case, the
eigenvalue problem cannot be reduced to a set of smaller (one-dimensional)
eigenvalue problems and one is left with a 2D or 3D problem.

In most cases, after discretization of the equations, the 
temporal stability problem
reduces to a generalized eigenvalue problem.
A method to solve such problems is to use the QZ algorithm.
Such an algorithm gives the full spectrum
of eigenvalues/eigenvectors, but the price
to pay for obtaining this
very rich information is very high in terms of memory requirement and 
CPU time consumption. 
Moreover the QZ algorithm does not
preserve the sparsity of the matrices.
On the other hand it is seldom needed to 
know the full spectrum: typically one is interested in the
few eigenvalues corresponding to the least stable or most unstable modes.
For the foregoing reasons it is important to be able to solve the
generalized eigenvalue problem with an iterative method which
preserves the sparsity of the matrices and converges
quickly and accurately to a small subset of the whole spectrum.
Three types of iterative methods exist to solve
such eigenproblems \cite{CHA.93,FOK.98}.
The first method is the subspace iterations method which is
just a generalization of the well-known power method. A second method is
the Jacobi-Davidson algorithm.
The third one is 
to use Krylov based methods such as the Arnoldi method or the unsymmetric
Lanczos method. Comparisons \cite{LEH.SCO.95} between the subspace
iterations method
and the Krylov type ones tend to show that the second ones are more efficient
when applied on large sparse matrices. We have chosen to use the Arnoldi method
because it is easy to implement. Its backward stability is now well understood
and it does not require any heuristics whereas numerical difficulties such as
serious breakdowns can be encountered using the unsymmetric Lanczos method.

In this paper we consider as a model problem the computation of the
inertial modes of a rotating spherical shell. This problem contains many
difficulties that make it a very good test case.  A first difficulty
is that the problem is essentially two-dimensional, because variables
such as the radial distance and the polar angle cannot be separated.
The size of the matrices thus grows very quickly with the resolution;
for parameters of physical interest matrices are of order $10^5$ or
larger. The second difficulty is that the
partial differential equations become of hyperbolic type and therefore
yields, with boundary conditions, an ill-posed problem \cite{RV97,RGV01}.
Solving this eigenvalue problem is therefore demanding numerically. 
The third difficulty is that matrices are highly non-normal.
The eigenvalue spectrum is thus very sensitive to machine precision
and special tools must be used to analyze and control the round-off errors.
Our analysis revealed some interesting aspects from the viewpoint
of numerical precision. In particular we think that our results on
the behaviour of round-off and spectral errors and their interplay
are useful in many fields of physics where two-dimensional eigenvalue
problems appear.

We organized the paper as follows: we first describe the physics of our
test problem and how we discretize it using spectral methods (section 2). We
also briefly recall the principle of the incomplete Arnoldi-Chebyshev
algorithm (section 3). We then discuss the role of spectral resolution
(sect. 4) and presents our results about the behaviour of round-off
errors (sect. 5); conclusions and outlooks follow.

\section{The test-problem}

\subsection{Formulation}

We consider the problem of finding
the modes of oscillation of a rigidly rotating fluid contained in a spherical
shell; it is investigated for its astrophysical and geophysical
applications (see \cite{RV97,RGV01} for a
detailed discussion). Eigenmodes of this system are called inertial
modes.

The fluid is contained between two spheres of radii $\eta R$ and $R$ ($\eta
<1$) and rotates at an angular velocity $\Omega$ around the $z$-axis. Choosing
$R$ as the length scale and $(2\Omega)^{-1}$ as the time scale, the
non-dimensional form of the equations for the equations governing
perturbations are:

\greq
E\Delta\na\times\vu - \na\times(\ez\times\vu) = \lambda\na\times\vu\\
\Div\vu=0
\egreqn{eqmo}
where $\lambda$ is the eigenvalue (a non-dimensional
frequency) and \mbox{$E=\nu/2\Omega R^2$} is the Ekman number. $E$
is the non-dimensional measure of the kinematic viscosity $\nu$ and is
usually a small parameter ($E<10^{-4}$).

Equations \eq{eqmo} are completed by boundary conditions on the velocity taken
at $r=\eta$ and $r=1$. We impose stress-free boundary conditions, namely
that

\[ u_r=\dr{}\lp\frac{u_\theta}{r}\rp=\dr{}\lp\frac{u_\varphi}{r}\rp=0\]
on the boundaries ($r,\theta,\varphi$ are the usual spherical
coordinates).

\subsection{Numerical method}

We discretize the preceding partial differential equations using
spectral methods because of their efficiency at convergence
\cite{fornberg,BGM99}.

 For obvious geometrical reasons, the angular part of the fields is
expanded on spherical harmonics; hence, we set

\[\vu=\sum_{l=0}^{+\infty}\sum_{m=-\ell}^{m=+\ell}\ulm(r)\RL+\vlm(r)\SL+\wlm(r)\TL
,\]

\noi where

\[\RL=\YL(\theta,\phi)\vec{e}_{r},\qquad \SL=\na\YL,\qquad \TL=\na\times\RL \]

\noi and where $\YL(\theta,\phi)$ are normalized spherical harmonics
(gradients in the definitions of $\SL$ and $\TL$ are taken on sphere of
unit radius).

Following some simple rules, given in \cite{rieu87b}, the equation of 
vorticity (\ref{eqmo}a)
may be projected rather easily on spherical harmonics.
The radial functions $\ulm(r)$ and $\wlm(r)$ then obey the following system

\beq\left\{ \begin{array}{l}
\biggl(E\Delta_\ell+\iml\biggl)\wlm + 
\disp{A_m(\ell)r^{\ell-1}\dnr{}\biggl(
\frac{u_m^{\ell-1}}{r^{\ell-2}}\biggr)  \\
\hspace{3.5cm}
\disp{+A_m(\ell+1)r^{-\ell-2}\dnr{}\biggl( r^{\ell+3}u_m^{\ell+1}\biggr)}
= \lambda \wlm }\\
\\
\biggl(E\Delta_l+\iml\biggl) \Delta_\ell(r\ulm) - 
\disp{B_m(\ell)r^{\ell-1}\dnr{}
\biggl(\frac{w_m^{\ell-1}}{r^{\ell-1}}\biggr) \\
\hspace{3cm}
\disp{-B_m(\ell+1)r^{-\ell-2} \dnr{}\biggl( r^{\ell+2}w_m^{\ell+1}\biggr)}
= \lambda \Delta_\ell(r\ulm) }
\end{array} \right. \eeqn{bigsys}

\noi where we have eliminated the $\vlm$'s using $\Div\vu=0$. The
following notations have also been introduced :

\[ A_m(\ell) = \frac{1}{\ell^2}\sqrt{\frac{\ell^2-m^2}{(2\ell\!-\!1)(2\ell\!+\!1)}}, \ 
B_m(\ell) = \ell^2(\ell^2-1)A_m(\ell), \  \Delta_\ell = \frac{1}{r}\ddnr{}r -
\frac{\ell(\ell+1)}{r^2} \]

\noi System \eq{bigsys} is an infinite set of differential equations
where the coupling between radial functions of indices $\ell\!-\!1$,
$\ell$ and $\ell\!+\!1$ is due to the Coriolis force. Note that different
$m$'s are not coupled.

For the discretization in radial coordinate we 
approximate the radial functions by truncated expansions of
$N\!+\!1$ Chebyshev polynomials. Thus, each of the functions may be represented
either by its spectral components or by its values on 
the Gauss-Lobatto collocation nodes. We use the latter
representation. In such case, differential operators $d^k/dr^k$ are
represented by full matrices of order ($N\!+\!1$). As system
\eq{bigsys} couples radial functions of indices $\ell-1$, $\ell$ and
$\ell+1$, it yields a generalized eigenvalue problem with tridiagonal
block matrices which we write symbolically:

\begin{equation}
\left( \begin{array}{cccc}
    \hat{A}^{\rm I}_{-1,\ell}  & \hat{A}^{\rm I}_{0,\ell}  & \hat{A}^{\rm
I}_{1,\ell} & 0 \\
0 & \hat{A}^{{\rm II}}_{-1,\ell} & \hat{A}^{{\rm II}}_{0,\ell} & \hat{A}^{{\rm
II}}_{1,\ell}    \\
\end{array}\right)
\left( \begin{array}{c}
\hat{u}_{\ell\!-\!1}\\
\hat{w}_{\ell}\\
\hat{u}_{\ell\!+\!1}\\
\hat{w}_{\ell\!+\!2}\\
\end{array}
\right)
=
\lambda
\left( \begin{array}{cccc}
    \hat{B}^{\rm I}_{-1,\ell}  & \hat{B}^{\rm I}_{0,\ell}  & \hat{B}^{\rm
I}_{1,\ell} & 0 \\
0 & \hat{B}^{{\rm II}}_{-1,\ell} & \hat{B}^{{\rm II}}_{0,\ell} & \hat{B}^{{\rm
II}}_{1,\ell}    \\
\end{array}\right)
\left( \begin{array}{c}
\hat{u}_{\ell\!-\!1}\\
\hat{w}_{\ell}\\
\hat{u}_{\ell\!+\!1}\\
\hat{w}_{\ell\!+\!2}\\
\end{array}
\right)
\label{eq:eig_prob}
\end{equation}

For each value of $\ell$,
$\hat{A}^{\rm I}_{-1,\ell}$, $\hat{A}^{\rm I}_{0,\ell}$ and $\hat{A}^{\rm
I}_{1,\ell}$,
are $(N\!+\!3)\times (N\!+\!3)$ matrices. 
They correspond to the discretization of the l.h.s. of
the first equation in (\ref{bigsys}) at the $N\!+\!1$ Gauss Lobatto 
nodes,
and of the boundary condition $r \frac{d \wlm}{dr} -\wlm=0$ imposed
at the two radial boundaries.
$\ell$ runs from
$m$ to $L$ by steps of two when $m$ is even and it runs from $m+1$ to $L$ 
by steps of two when $m$ is odd. 
Similarly $\hat{A}^{\rm II}_{-1,\ell}$, $\hat{A}^{\rm II}_{0,\ell}$ and
$\hat{A}^{\rm II}_{1,\ell}$
are $(N\!+\!5)\times (N\!+\!5)$ matrices corresponding to the
discretization of the l.h.s. of the second equation in (\ref{bigsys})
at the Gauss Lobatto nodes, plus boundary conditions $\ulm=\frac{d^2
\ulm}{dr^2}+\frac{2}{r}\frac{d\ulm}{dr}=0$ at the two radial boundaries.

\section{The Incomplete Arnoldi-Chebyshev Method}

For efficiency reasons and memory requirements, the 
generalized eigenvalue problem \eq{eq:eig_prob}
should be solved using an iterative method
because the matrices are large and sparse. As previously stated, 
a good method is
the incomplete Arnoldi-Chebyshev method { which we now briefly
recall}. 

Let $K(u,A) = \{ u, Au, \dots, A^{m-1}u \}$ be the Krylov subspace built
from the initial vector
$u$, $V_m=\{v_i\}_{i=1 \ldots m}$ of size $n \times m$ be an orthonormal
basis of this subspace.

For applications to stability problems, one is mostly 
interested in the least-stable (or most unstable) eigenmodes which
are associated with the generalized eigenvalues $\lambda$
with the greatest real part.
Since these eigenvalues obviously
do not belong to the outside part of the spectrum, we have to perform a
spectral transformation.
Let $(\mu,\overline{y})$ be the solutions of

\beq
([A]-\sigma [B])^{-1} [B] \overline{y} = \mu \overline{y}.
\label{eq:sigma}
\eeq

Then, one easily shows that $(\lambda=\sigma+1/\mu,\; \overline{x}=
\overline{y})$. Thanks to this spectral transformation, the eigenvalues near
the shift (the guess)
$\sigma$ are sent to the outside part of the spectrum and the Arnoldi method
can now deliver the desired eigenpair very efficiently. 
The derived method can be summarized as
follows
\begin{algorithm}
Parameter: integers $r$ (number of desired eigenpairs), $m$ (number of Arnoldi
steps), with $r \le m \ll n$, Arnoldi starting vector $u$ and
degree $k$ of Chebyshev acceleration polynomial.

$1$. Perform $m$ steps of the Arnoldi method starting from $u$ to compute
$V_m$ and $H_m$:
$$ ([A]-\sigma [B])^{-1} [B] V_m = V_m H_m +v_{m+1} e_m^T.$$

$2$. Compute the eigenpairs $(\mu_i,y_i)_{i=1 \colon m}$ by applying the QR
algorithm to $H_m$:
$$H_my_i=\mu_i y_i.$$

$3$. If the stopping criterion is satisfied for the $r$ wanted eigenvalues
then go to step $6$.

$4$. Compute the parameters of the ellipse containing the $m-r$ unwanted
eigenvalues of $H_m$ and set $z_0=\sum_{i=1}^r {\alpha_i V_m y_i}$
where
$$\alpha_i=\frac{\norm{([A]-\sigma [B])^{-1} [B])V_my_i-\mu_i V_my_i}}
{(\norm{[A]-\sigma [B])^{-1} [B]}\norm{V_my_i}}.$$

$5$. Perform $k$ steps of the Chebyshev acceleration starting from $z_0$
to obtain a better starting vector $u$ for the Arnoldi method; go to step $1$.

$6$. Set $\{\lambda_i = \sigma+1/\mu_i, \overline{x_i}=V_m
y_i\}_{i=1\colon r}$.

\end{algorithm}

This algorithm requires a matrix-vector product involving the matrix
$[B]$ and a linear solver to compute $z_2$ solution of $([A] - \sigma [B])
z_2 = z_1$.
In our application, $[A]$ and $[B]$ are banded matrices, so
that a band linear solver from LAPACK has been used.
The internal dense eigensolver in step 2 has been taken from EISPACK.
The interested reader is referred to
\cite{BRA.94b,BEBRADU94,BRACHADUN94,BRA.95} for more details.

\section{The role of spatial resolution}

We first study the convergence of eigenvalues and eigenvectors
as a function of the resolution.
The two relevant parameters are the 
degree of the largest Chebyshev polynomial (equal to the 
number of radial nodes $N$ minus one) and 
the degree $L$ of the last spherical harmonic.
We deal only with axisymmetric $m\!=\!0$ modes
and therefore drop the index $m$ (no additional difficulty arises when
$m\neq0$).

In the following we use the notation $\omega$ for
the imaginary part of the eigenvalue (the frequency), and 
$\tau$ for the real part. Thus $\lambda \equiv \tau + i \omega$.
All the modes of our test-problem are stable, i.e. $\tau<0$, and
$|\tau|$ is the damping rate.

We define the Chebyshev and Legendre spectra of the field $u$
with spectral components $u(\ell,n)$ in the following way:

$${\cal C}(n)={\max_\ell |u(\ell,n)|\over \max_{\ell,n} |u(\ell,n)|}
\qquad \quad
{\cal L}(\ell)={\max_n |u(\ell,n)|\over \max_{\ell,n} |u(\ell,n)|}
$$

Both spectra are filled because inertial modes display
very fine structures (see \cite{RV97}
for 
typical spectra and eigenfunctions occurring at Ekman numbers
as low as $E=10^{-8}$).

\begin{figure}
   \subfigure[]{
      \label{fig:e1e-4.re-.0}
      \begin{minipage}[b]{0.5\linewidth}
         \centering \includegraphics[width=0.9\linewidth]{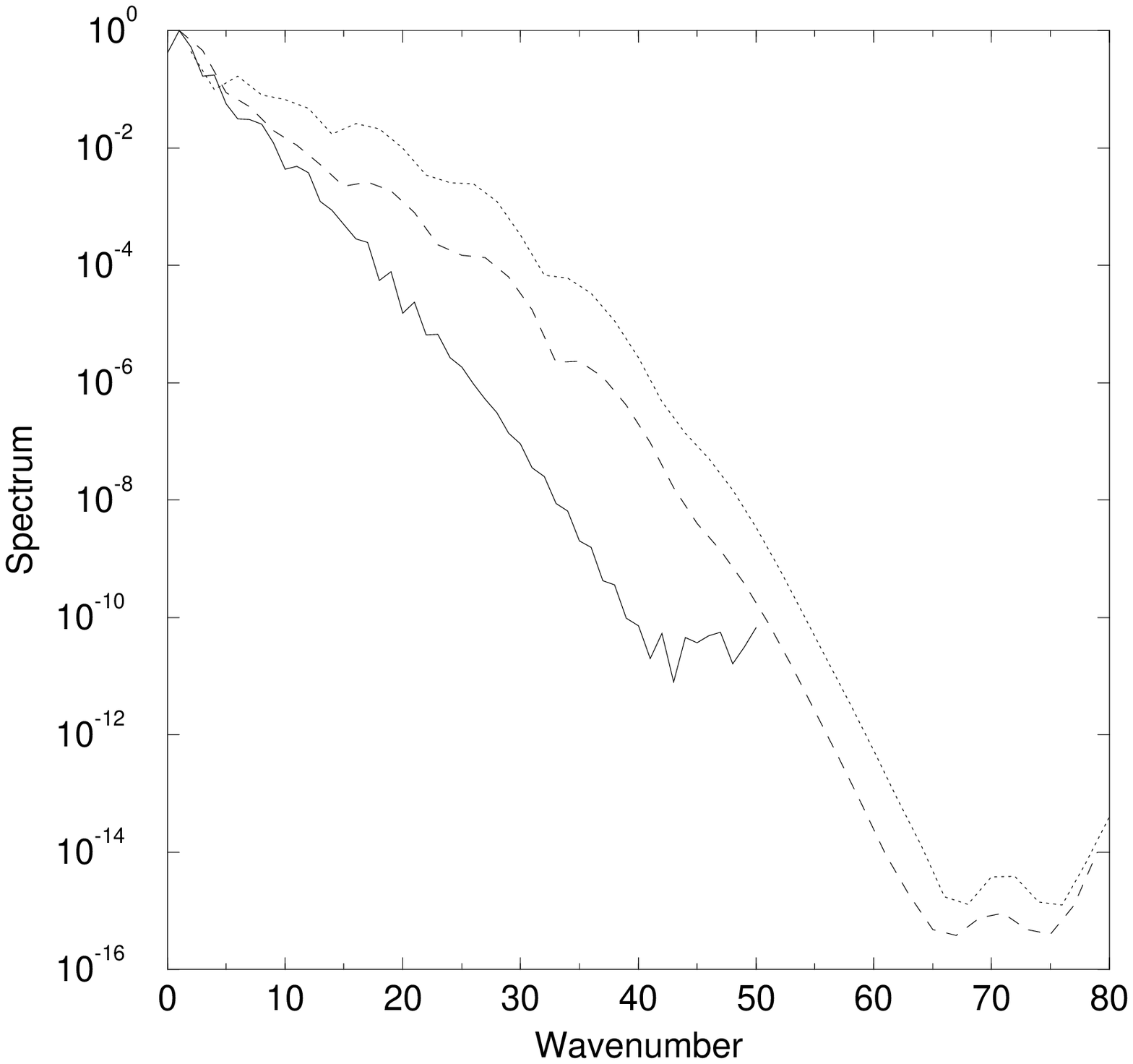}
      \end{minipage}
   }
   \hfill
   \subfigure[]{
      \label{fig:e1e-4.re-.5}
      \begin{minipage}[b]{0.5\linewidth}
         \centering \includegraphics[width=0.9\linewidth]{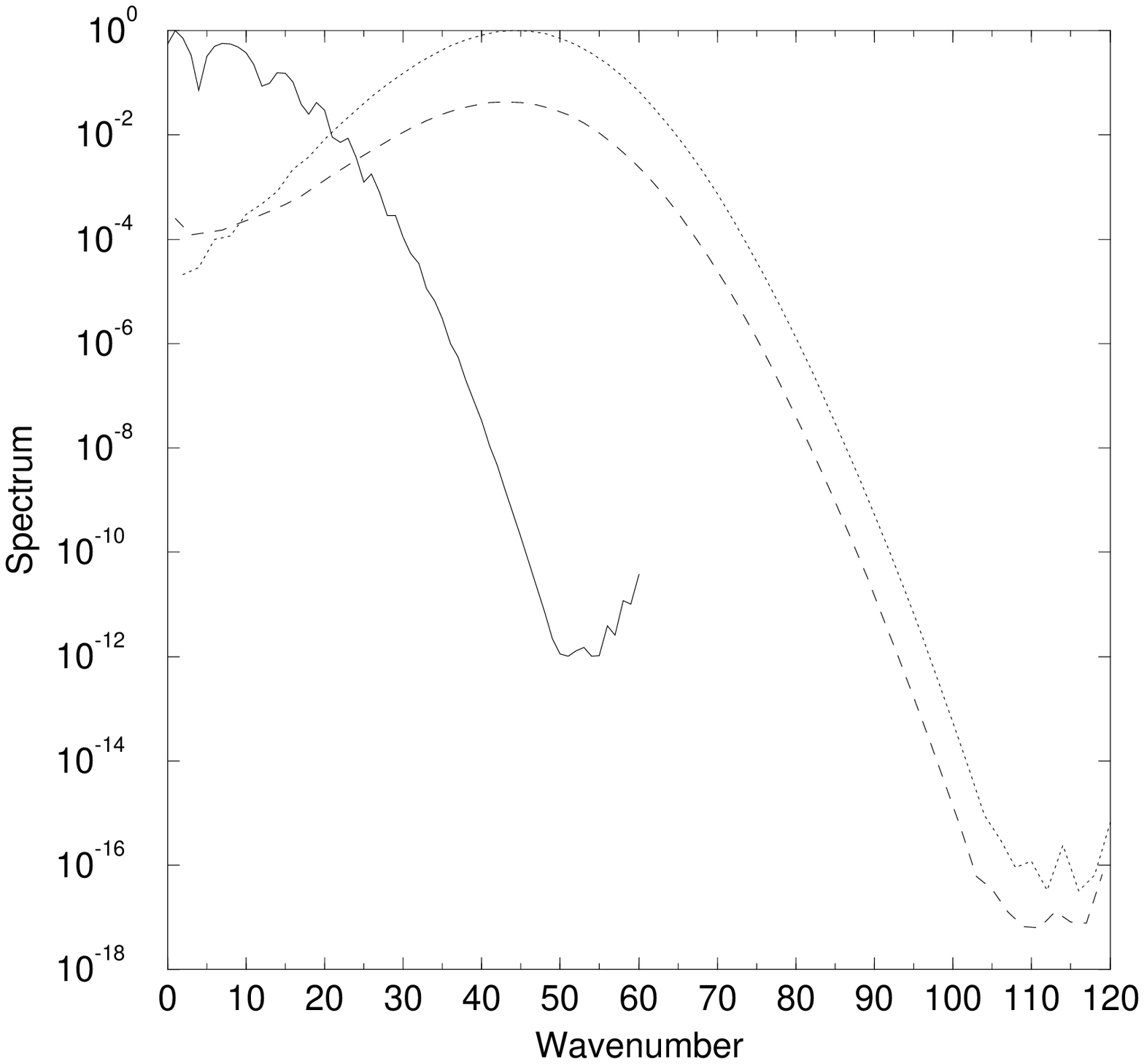}
      \end{minipage}
   }
   \caption{
Chebyshev (solid line) and Legendre (dashed
line for $ru$ and dotted line for $w$) spectra.
(a): mode at $E=10^{-4}$ with $\omega=0.657976$
and $\tau=-0.00875$.
(b): mode at $E=10^{-4}$ with $\omega=0.654580$
and $\tau=-0.51$.
}
\end{figure}

Here, we take a moderately small value of the Ekman number: $E=10^{-4}$ so that
the full eigenvalue spectrum can be explored with an affordable resolution.
As it may be expected, eigenvalues with smaller
$|\tau|$ require less
resolution to converge than those with large damping rate
(see figures \ref{fig:e1e-4.re-.0} and \ref{fig:e1e-4.re-.5}).
This is easily
understood since eigenvectors with small $|\tau|$ tend to 
have a smoother pattern, which is well approximated
by a small number of spectral modes.

      \psfrag{re0}{\tiny $\tau\!=\!0$}
      \psfrag{re-.0}{\tiny $\tau\!=\!0$}
      \psfrag{re-.20}{\tiny $\tau\!=\!-.20$}
      \psfrag{re-.3}{\tiny $\tau\!=\!-.3$}
      \psfrag{re-.39}{\tiny $\tau\!=\!-.39$}
      \psfrag{re-.4}{\tiny $\tau\!=\!-.4$}
      \psfrag{re-.43}{\tiny $\tau\!=\!-.43$}
      \psfrag{re-.5}{\tiny $\tau\!=\!-.5$}
      \psfrag{re-.51}{\tiny $\tau\!=\!-.51$}

The convergence of eigenvalues 
as a function of spatial resolution goes together 
with that of eigenvectors: unless all the scales present in
the eigenvector are resolved, both the eigenvector and the
eigenvalue are not well approximated. 
This observation allows us to give a simple rule
to check the convergence of the eigenvalue.
Let us define the ratio $f_L$ between the
lowest spherical harmonics coefficient
and the largest one,
and define $g_N$ as the same ratio but for the Chebyshev expansion.
\beq
f_L=\frac{\min {\cal L}(l)}{ \max {\cal L}(l)}\qquad\quad
g_N=\frac{\min {\cal C}(n)}{ \max {\cal C}(n)}
\label{eq:flgn}
\eeq
These two ratios measure the truncation error in the spherical harmonic
expansion ($f_L$) and in the Chebyshev expansion ($g_N$).  We next
define $\varepsilon$ as the absolute value of the difference between
the computed eigenvalue and the { converged one (i.e. obtained with a
large resolution).}

\begin{figure}
   \begin{minipage}[t]{0.48\linewidth}
      \centering \includegraphics[width=1.0\linewidth]{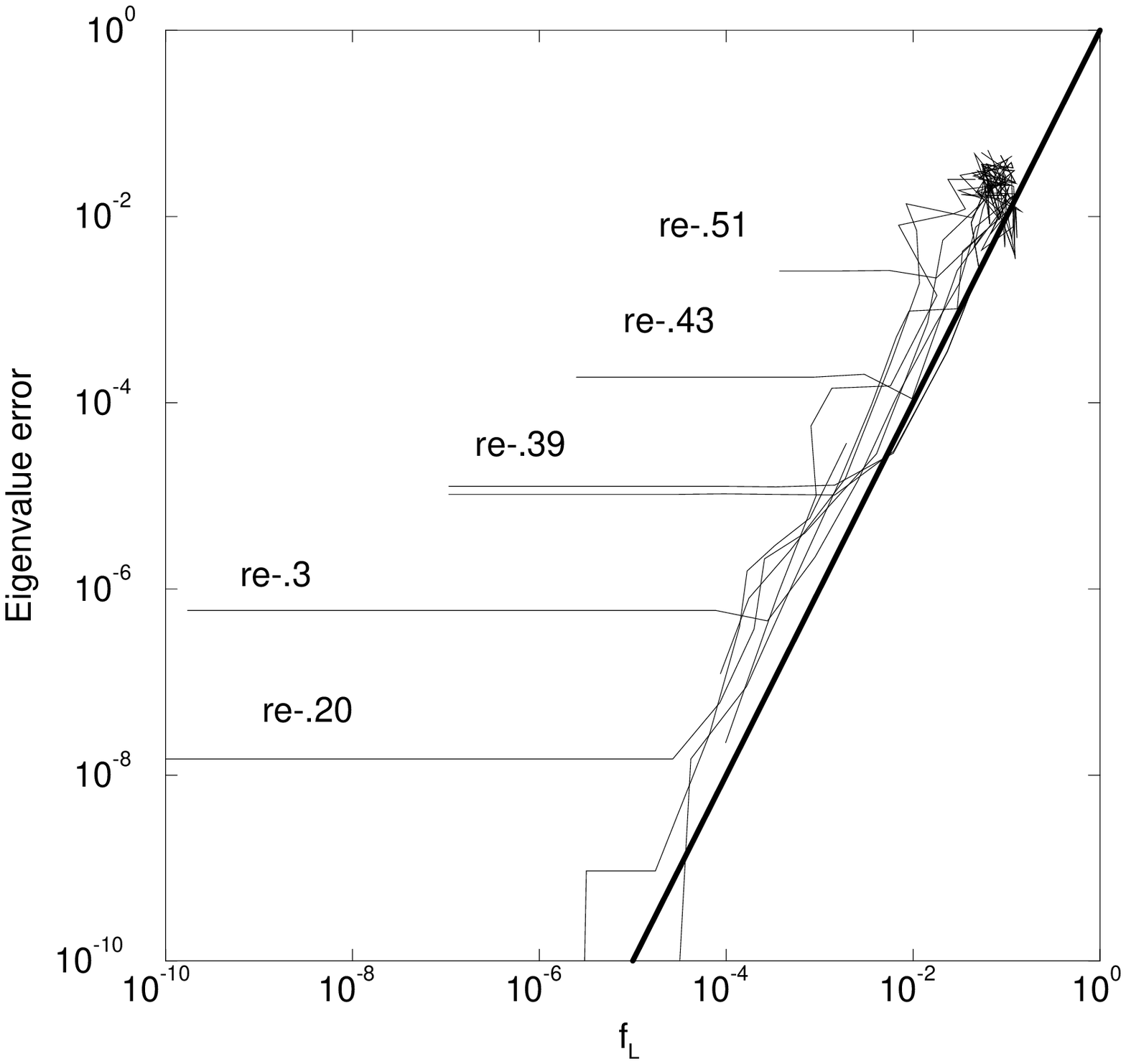}
      \caption{Error $\epsilon$ of the computed eigenvalue
plotted as a function of the Legendre truncation error of the eigenvector
$f_L$ (eq. (\ref{eq:flgn})).
Different curves correspond to 
different eigenmodes.
Thick line corresponds to the law $\varepsilon= f_L^2$.
Note the horizontal plateau at large resolutions, due to round off errors.
}
      \label{fig:err_leg}
   \end{minipage}
   \hfill
   \begin{minipage}[t]{0.48\linewidth}
      \centering \includegraphics[width=1.0\linewidth]{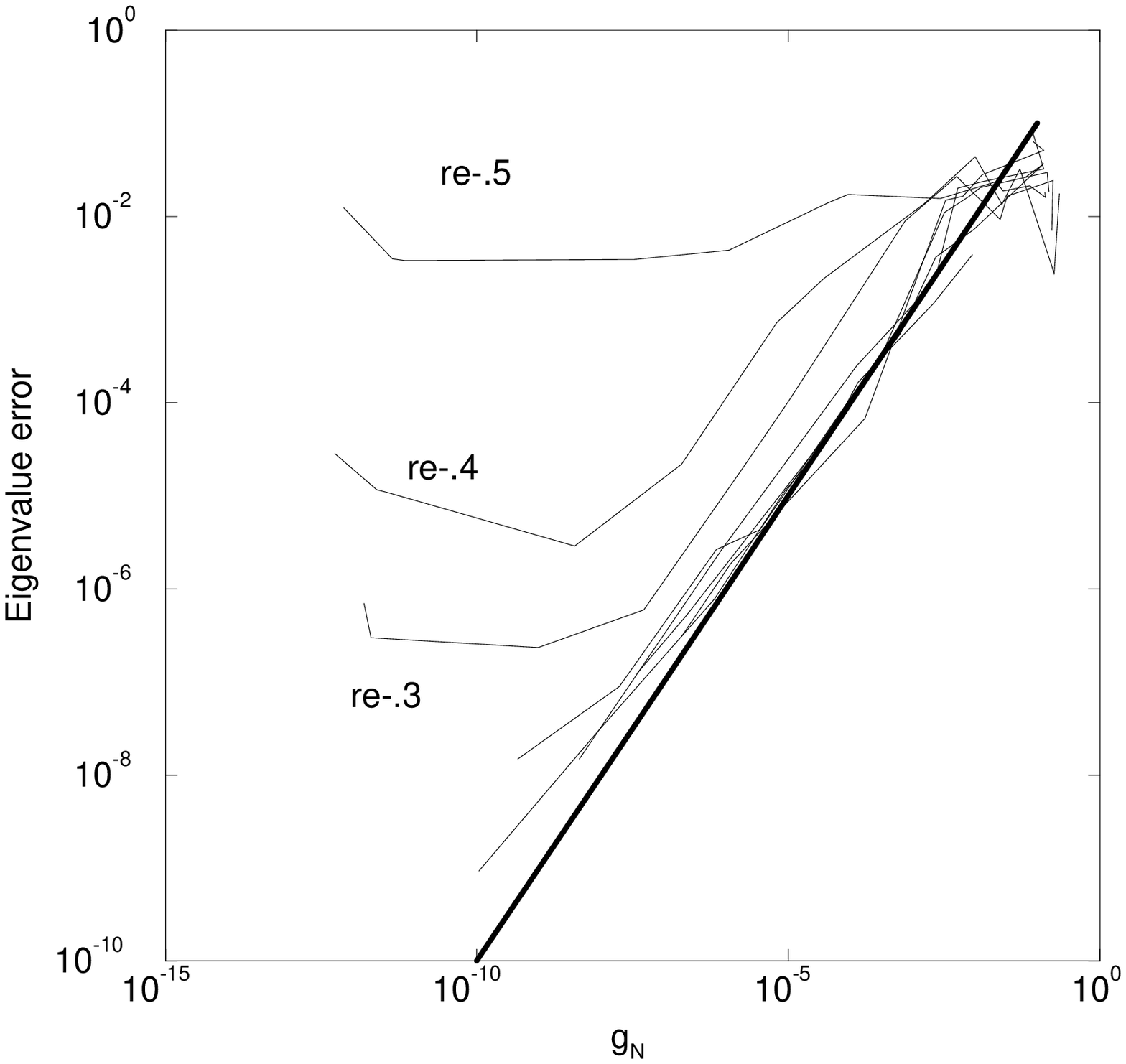}
      \caption{Error $\epsilon$ of the computed eigenvalue  
plotted as a function of the Chebyshev truncation
error of the eigenvector
$g_N$ (eq. (\ref{eq:flgn})). Thick line corresponds
to the law $\varepsilon= g_N$.}
      \label{fig:err_cheb}
   \end{minipage}
\end{figure}

In figure \ref{fig:err_leg} we plot $\varepsilon$ as a function of $f_L$.
The number of Chebyshev polynomials was chosen large enough to resolve
completely the radial dependence.  We clearly see that $\varepsilon$
follows the law $\varepsilon \propto f_L^2$ until a plateau is reached.
The plateau appears at the largest resolutions and indicates that no
better approximation to the eigenvalue can be obtained by increasing
the resolution. It gives thus a measure of the round-off error of
the computation. { From the curves obtained for different eigenmodes,}
we see that {\em the round-off error is a rapidly increasing function
of the damping rate.}

In figure \ref{fig:err_cheb} we plot $\varepsilon$ as a function of the
parameter $g_N$.  Here the number of spherical harmonics was set large
enough to fully resolve the angular dependence.  The Chebyshev convergence
appears to be governed by the law $\varepsilon \propto g_N$. Here too,
good convergence is obtained only for least-damped modes. We note also
that the plateau values are very close to those of the preceding figure.

\section{The importance of round-off errors}

The foregoing results indicate that round-off errors play a major role
in the accuracy of the numerical solution, especially for strongly
damped modes. We shall now investigate this point more thoroughly.

\begin{figure}
   \centering \includegraphics[width=0.48\linewidth]{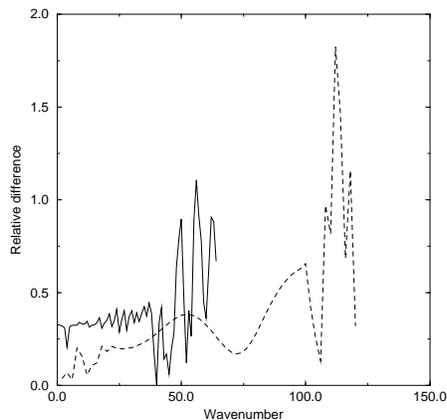}
   \caption{
Relative difference of the spectral coefficients obtained with two 
computations where the only difference
is a change of the Arnoldi-Chebyshev shift $\sigma$ of the Arnoldi-Chebyshev
algorithm. The mode is that of figure~\ref{fig:e1e-4.re-.5}.
Dashed line: Legendre coefficients
${\delta\cal L}(n)$. Solid line: Chebyshev coefficients
${\delta\cal C}(n)$.
}
   \label{fig:erreur_e1e-4.re-.5}
\end{figure}

First, we stress the fact that good spectral convergence is not at
all a guarantee against round-off errors. This point can be made very
clear using the mode displayed on figure~\ref{fig:e1e-4.re-.5}
for example. No doubts that for such a mode the spectral expansion has
converged: there are 12 decades
in the Chebyshev spectrum and 16 decades in the Legendre spectrum;
however the whole spectrum is subject
to large round-off error at all wavenumbers. 
To illustrate this point we consider two different computations where 
we only change the value of the 
shift $\sigma$ (see equation (\ref{eq:sigma}))
of the Arnoldi-Chebyshev
algorithm: $\sigma=-0.51 + i 0.65458$ in the first case and 
$\sigma=-0.51 + i 0.65558$ in the second case. 
In both computations
the Ekman number is $E=10^{-4},$ $L=120$ and $N=64$;
the Arnoldi-Chebyshev algorithm converges to the same eigenmode.
The Chebyshev and Legendre spectra for the first case are those
represented on figure~\ref{fig:e1e-4.re-.5}; the two spectra for 
the second case are similar.
We plot in figure \ref{fig:erreur_e1e-4.re-.5}
the relative difference of the spectral coefficients, defined as 
$$
{\delta\cal C}(n)={|{\cal C}_2(n)-{\cal C}_1(n)|\over0.5( {\cal
C}_2(n)+{\cal C}_1(n))},\qquad \quad
{\delta\cal L}(n)={|{\cal L}_2(n)-{\cal L}_1(n)|\over0.5( {\cal
L}_2(n)+{\cal L}_1(n))}
$$
where subscript $1$ (resp. $2$) corresponds to first (resp. second)
eigenvector.
We see that the relative difference is spread almost
uniformly throughout the wavenumbers,
until round-off error in the spectrum is reached, where necessarily
the relative error grows to ${\cal O}(1)$.
This uniform spreading is not surprising; in 
\cite{arioli95} it is shown that 
for Chebyshev expansions the spectral 
round-off error of differential operators
is distributed uniformly among wavenumbers.

The round-off error 
may be investigated quite systematically by computing the spectral portrait of
this eigenvalue problem.

Spectral portraits and pseudospectra have recently attracted the attention
as a tool of choice for investigating spectral
properties of nonnormal matrices (see \cite{CHA.FRA.96,TOU.96,TRE.TRE.RED.DRI.93}).
It consists in the representation of the map  
\[z \longrightarrow
{\rm spp}(z)=\log_{10}\left[ \Vert{(A-zB)^{-1}}\Vert_2 
(\Vert A \Vert_2 + \vert z \vert \Vert B \Vert_2)\right] \]
in a prescribed region of the complex plane.
The contour lines of level $\varepsilon$ of the spectral portrait are the
borders of the $\varepsilon$-pseudospectrum of the matrix pair $(A,B)$~:
they enclose all the eigenvalues of the matrix pairs $(A+\Delta A, B +
\Delta B)$ and with $\Vert \Delta A \Vert_2 \leq \varepsilon 
\Vert A \Vert_2$ and 
$\Vert \Delta B \Vert_2 \leq \varepsilon \Vert B \Vert_2$.

If $\varepsilon$ is chosen as the backward error for a computed eigenvalue
$\tilde{\lambda}$, then the contour line of level $\varepsilon$ encloses
all the complex numbers with the same backward error $\varepsilon$ for the
pair $(A,B)$. The larger the enclosed area, the worse-conditioned the
eigenvalue. The diameter of the enclosed area gives an idea of the largest
possible relative error on $\tilde{\lambda}$.
For a semi-simple eigenvalue, it is always possible to bound the error on
$\tilde{\lambda}$ by the product of the condition number and the backward
error. This is not possible for multiple defective eigenvalues and the
spectral portrait is a useful alternative.

On the example studied here, the computation is backward stable : we can
then look at the contour line of level machine precision that is
$10^{-16}$ (only ``$-\log_{10} \varepsilon$'' appears on the figures). 
We see that this level curve encloses a large region of the spectrum, which
tends to indicate a significant spectral instability in the matrix pairs
under study.

\begin{figure}
      \centering \includegraphics[width=0.6\linewidth]{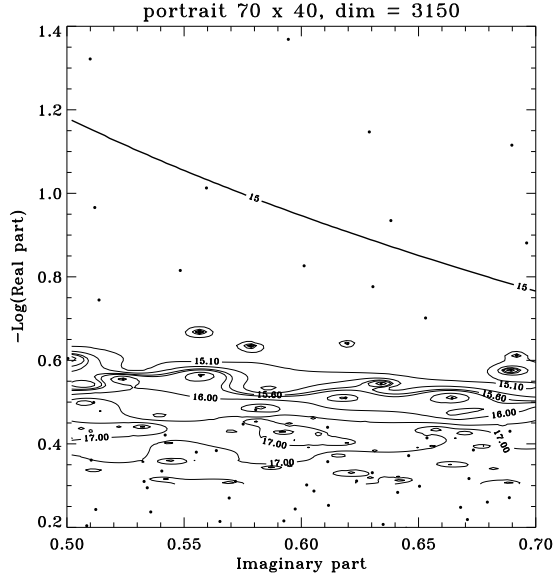}
      \caption{Spectral portrait. $E=10^{-4}$, $L=70$, $N=40$.}
      \label{fig:portrait.ek1e-4.L70.Nr40}
\end{figure}

We display in figure \ref{fig:portrait.ek1e-4.L70.Nr40} the spectral portrait 
for { our eigenvalue problem using a resolution of $L=70$ and $N=40$
which corresponds to matrices of order 3150}.
We superpose 
the eigenvalues obtained using the QZ algorithm (black points)
and the isolines of spectral portrait. For values of the 
spectral portrait larger than approximately 16 { (lower part of the
figure) }the 
computed eigenvalues are completely undetermined in double precision. 
This corresponds to damping rates larger than 0.25 approximately. 

\begin{figure}
   \begin{minipage}[t]{0.48\linewidth}
   \centering \includegraphics[width=1.0\linewidth]{fig/valp_bruit_matpert_5000.delta.eps}
   \caption{Plot of 
several eigenvalues
obtained by perturbing randomly the two matrices $A$ and $B$ of
eq. (\ref{eq:sigma}). The magnitude of the perturbation is
the machine precision $2.22\times 10^{-16}$.
The shift is a fixed value near the the exact eigenvalue.
The Ekman number is $E=10^{-4}$ and the resolution
$L=94$, $N=50$. Each black dot in the plot is the difference between the 
computed eigenvalue and the exact one 
$\lambda=-0.38521\,09005\,33277\!+\!i 0.65359\,27894\,40845$.
}
   \label{fig:bruit matpert}
   \end{minipage}
   \hfill
   \begin{minipage}[t]{0.48\linewidth}
   \centering \includegraphics[width=1.0\linewidth]{fig/valp_bruit_shift_5000.delta.eps}
   \caption{Plot of 
several eigenvalues obtained
by making different calculations where the only change 
is the shift parameter $\sigma$ of eq. (\ref{eq:sigma}).
$\sigma$ is changed by a random perturbation 
of magnitude $10^{-5}$
near the exact eigenvalue.
The parameters are the same as in figure \ref{fig:bruit matpert}.
We remark that the two plots are hardly distinguishable.
}
   \label{fig:bruit shift}
   \end{minipage}
   \hfill
\end{figure}

 \begin{figure}
    \centering \includegraphics[width=0.48\linewidth]{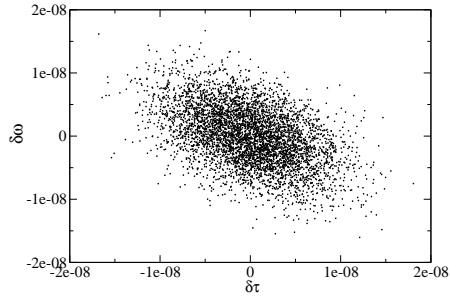}
    \caption{
Same as figure \ref{fig:bruit matpert}, only the calculations are
done using extended precision.
}
   \label{fig:bruit matpert q}
 \end{figure}

However, computation of pseudospectra is an expensive task and it is
therefore not feasible on production runs.  There have been recent
developments in the algorithms whereby one can obtain an approximation
to the pseudospectra in a region near the interesting eigenvalues at
reasonable cost \cite{TRE.99,WRI.01}. However, those techniques must be used
with special care as they are not totally reliable in the case where the
matrix is highly nonnormal.  A cheap technique that is used routinely to
evaluate the sensitivity of eigenvalues to round-off error is to compute
the eigenvalues of randomly perturbed matrices.  This technique can be
used without further coding on any eigenvalue solver, yet one must code
the perturbations to the matrix elements.  In the following we
explore the impact of round-off errors by means of matrix perturbation,
and present a new technique which gives the same results but
does not require any coding at all.

\begin{figure}
   \centering \includegraphics[width=0.8\linewidth]{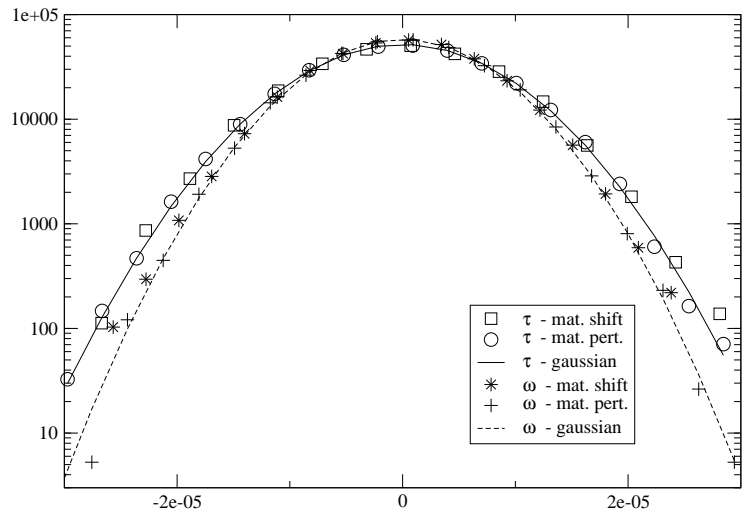}
   \caption{Probability density functions for the computed eigenvalues of figures 
\ref{fig:bruit matpert} and \ref{fig:bruit shift}. In abscissas 
are the differences between the real part of the eigenvalue
and the average value $\overline{\tau}$ for the
preturbed shift case (squares) and the perturbed matrix case (circles),
and the differences between the imaginary part of the eigenvalue
and the average value $\overline{\omega}$ for the
preturbed shift case (stars) and the perturbed matrix case (plus).
The continuous and broken lines corresponds to the gaussian curves which 
fita at best the data. We see that gaussian fit is almost perfect.
}
   \label{fig:hist}
\end{figure}

In figure \ref{fig:bruit matpert} we plot the eigenvalues obtained by
making several calculations on perturbed matrices.  Each point in the
figure is the eigenvalue obtained by perturbing the two matrices $A$
and $B$ of eq. (\ref{eq:sigma}) by random values uniformly distributed in
the interval  $(-\epsilon_m, \epsilon_m)$, where $\epsilon_m=2.22\times
10^{-16}$ is the machine precision.  The eigenvalues form a cloud
of points concentrated in the neighbourhood of the exact eigenvalue
$\lambda=-0.38521\,09005\,33277\!+\!i 0.65359\,27894\,40845$ (this
eigenvalue has been obtained using extended precision; it is the ``exact"
eigenvalue of the truncated problem (\ref{eq:eig_prob}) and not the one of
the differential problem (\ref{bigsys})).  We did a statistical analysis
on a large number of eigenvalues (50000). The results are summarized in
table \ref{table:statistics} and figure \ref{fig:hist}.  Both the real
and imaginary parts follow quite well a Gaussian law: this can be seen in
figure \ref{fig:hist} where the probability density functions of computed
eigenvalues are plotted together with the gaussian curve that fits at
best the data. More quantitatively, the skewness and kurtosis (table
\ref{table:statistics}) are very close to those of the normal distribution
(resp. 0 and 3).  The order of magnitude of the round-off error is given
by the standard deviation of the data which is $\sigma_\tau\simeq
7.69\times 10^{-6}$ for the real part and $\sigma_\omega\simeq
6.83\times 10^{-6}$ for the imaginary part.  Note that the covariance
$\sigma_{\tau\omega}=\sum_{i=1}^n(\tau_i-\overline{\tau})(\omega_i-\overline{\omega})$
(like the correlation coefficient
$\rho_{\tau\omega}=\frac{\sigma_{\tau\omega}}{\sigma_\tau\sigma_\omega}$)
is small but non-zero, which means that the error distributions for the
real and imaginary parts are slightly  correlated.  We remark that the
standard deviations $\sigma_\tau$ and $\sigma_\omega$ have similar values,
i.e. the round-off error on $\tau$ is of the same order of magnitude as
that on $\omega$, even though $|\tau| << \omega$.

The standard deviations $\sigma_\tau$ and $\sigma_\omega$ turn out to
be essentially independent of the number of Chebyshev polynomials and
spherical harmonics, provided that both spectra are well resolved. The
values increase when the damping rate of the mode is increased, in
perfect accordance with the plateaux observed in figures~\ref{fig:err_leg}
and \ref{fig:err_cheb}.

\begin{table}
\begin{center}
\begin{tabular}{l|c|c|c|}

 &\hpp Matrix perturbation\hpp &\hpp Shift perturbation\hpp &\hpp Matrix
perturbation in quad. prec.\hpp \\ \hline
$\overline{\tau}$ & $-0.38520\,966$ & $-0.38521\,089$ & $-0.38521\,09005\,23$\\ \hline
$\overline{\omega}$& 0.65359\,249& $0.65359\,283$ &  0.65359\,27894\,14\\ \hline
$\overline{\tau}-\tau_{QP}$   &$1.24\times 10^{-6}$ & $8.67\times 10^{-9}$& $1.04\times 10^{-11}$ \\ \hline
$\overline{\omega}-\omega_{QP}$ &$-2.95\times 10^{-7}$ & $3.67\times 10^{-8}$& $-2.65\times 10^{-11}$\\ \hline
$\sigma_\tau$ & $7.69\times 10^{-6}$ & $7.80 \times 10^{-6}$& $4.99\times 10^{-9}$  \\ \hline
$\sigma_\omega$ & $6.83\times 10^{-6}$ & $6.91 \times 10^{-6}$ & $4.68\times 10^{-9}$ \\ \hline
$\rho_{\tau\omega}$ & $0.173$ & $0.142$ & $-0.476$\\ \hline
skewness($\tau$)   & $-0.016$ & $-0.004$ & 0.004\\ \hline
skewness($\omega$) & $-0.003$ & $-0.011$ & 0.0004\\ \hline
kurtosis($\tau$)   & $2.96$ & $2.98$ & $2.93$\\ \hline
kurtosis($\omega$) & $2.98$ & $2.98$ & $2.95$\\ \hline
\end{tabular}
\vspace*{2pt}
\caption[]{
Statistics for the computed eigenvalues of figures
\ref{fig:bruit matpert} and \ref{fig:bruit shift}.
We give the values of the averages $\overline{\tau}$
and $\overline{\omega}$, standard deviations $\sigma_\tau$
and $\sigma_\omega$, cross correlation $\rho_{\tau\omega}$,
skewness and kurtosis for the perturbed matrix case 
(column 2), the perturbed shift case (column 3) and
the perturbed matrix case using extended precision.\\
$(\tau_{QP}\!=\!-0.38521\,09005\,33277,\ \omega_{QP}\!=\!0.65359\,27894\,40845)$ 
stands for the ``exact"  eigenvalue computed with
quadruple precision
}
\label{table:statistics}
\end{center}
\end{table}

In a second series of 50000 computations we did not perturb the matrices
$A$ and $B$ but instead we perturbed the value of the Arnoldi-Chebyshev
shift $\sigma$ by a small random quantity near the exact eigenvalue.  With
this method there is no need to modify the source code for the eigenvalue
solver and/or for the construction of the matrices: we only need to
change the shift parameter $\sigma$ on input to the eigenvalue solver.
We obtain a cloud of eigenvalues (figure \ref{fig:bruit shift}) which
looks almost identical to that obtained in figure \ref{fig:bruit matpert}.
Each point in the figure is the eigenvalue obtained by changing the real
and imaginary part of the shift around the exact eigenvalue by random
values uniformly distributed in the interval  $(-10^{-5}, 10^{-5})$.
Actually we have verified that the statistics does not depend on the
amplitude of the shift perturbation.  So there is no need to know a
priori the exact value of the eigenvalue: any value of the shift which
delivers the wanted eigenmode is good.  The statistical values in table
\ref{table:statistics} confirm that the statistics of the eigenvalues
obtained in the two approaches are almost the same.

In a third series of 50000 computations the matrices $A$ and $B$ are
perturbed as in the first series by random values uniformly distributed in
the interval  $(-\epsilon_m, \epsilon_m)$, where $\epsilon_m=2.22\times
10^{-16}$.  However in this series the computation is performed using
extended precision.  Thus we measure directly the sensitivity of the
eigenvalues to perturbation of the matrices; in other words we compute
the $\epsilon_m$-pseudoeigenvalue.  From table \ref{table:statistics}
we see that there are about three digits of difference between the
standard deviations of the first two series and the present one: this
means that the Arnoldi-Chebyshev algorithm adds an extra factor of order
of magnitude $10^3$ to the round-off error.

\section{Conclusions}

We have analyzed in this paper the errors
that arise from the discretization and numerical computation
of partial differential eigenvalue problems yielding large matrices.
We have chosen as a model problem the two-dimensional eigenvalue problem
yielded by the computation of inertial modes in a spherical shell.

We have solved this problem using spectral methods for discretization
and the incomplete Arnoldi-Chebyshev algorithm for solving the eigenvalue
problem. The combination of these methods provides an efficient solver
for these large (two-dimensional) eigenvalue problems.

We have shown that the convergence of the eigenvalue and the
eigenvector, with respect to spatial truncation,
are tightly related: the absolute error of the eigenvalue decreases
linearly with the Chebyshev truncation error, and quadratically
with the spherical harmonics truncation error, until round-off error
becomes dominant.

We found that most-damped modes are the most ill-conditioned and are
therefore more sensitive to round-off error. This is made clear by the
spectral portrait of the linear operator. Its computation is however
very expensive and can be done only on small test problems.  We have
shown that a good estimation of the round-off error can be done in
practice by performing different computations changing only the value
of the Arnoldi-Chebyshev shift parameter on input; there is no need
to do extra coding and/or to use external tools. It turns out that the
round-off error on eigenvalues has an almost normal distribution, a result
which can be used to reduce this kind of error.  If the computation of
a single eigenmode is not too expensive one could take advantage of
this distribution of errors and perform $N$ computations with random
shifts; one can thus reduce the round-off error of the estimated
eigenvalue by a factor $\sqrt{N}$.

\bibliography{astroeig}

\clearpage
\end{document}